\documentclass{article}
\usepackage{waspaa23,amsmath,graphicx,url,times}
\usepackage{color}
\usepackage{bm}
\usepackage{graphicx}
\usepackage{cite}
\usepackage{amsmath,amssymb,amsfonts,mathrsfs}
\usepackage{multirow}
\usepackage{graphicx}
\usepackage{amsxtra}
\usepackage{threeparttable}
\usepackage{cite}
\usepackage{mathrsfs}
\usepackage{algorithm,algorithmic}
\usepackage{multirow, booktabs}
\usepackage{graphicx}
\usepackage{textcomp}
\usepackage{xcolor}
\usepackage{url}
\usepackage{hyperref}
\PassOptionsToPackage{hyphens}{url}

\DeclareMathOperator*{\argmin}{argmin}

\title{Signal Reconstruction from Mel-spectrogram Based on \\
Bi-level Consistency of Full-band Magnitude and Phase}

\name{Yoshiki Masuyama,$^{1}$
      Natsuki Ueno,$^{1}$
      Nobutaka Ono,$^{1}$}
\address{\vspace{-4pt} $^1$Tokyo Metropolitan University, Japan}

\begin{document}

\ninept
\maketitle

\begin{sloppy}
\begin{abstract}
We propose an optimization-based method for reconstructing a time-domain signal from a low-dimensional spectral representation such as a mel-spectrogram.
Phase reconstruction has been studied to reconstruct a time-domain signal from the full-band short-time Fourier transform (STFT) magnitude.
The Griffin--Lim algorithm (GLA) has been widely used because it relies only on the redundancy of STFT and is applicable to various audio signals.
In this paper, we jointly reconstruct the full-band magnitude and phase by considering the bi-level relationships among the time-domain signal, its STFT coefficients, and its mel-spectrogram.
The proposed method is formulated as a rigorous optimization problem and estimates the full-band magnitude based on the criterion used in GLA.
Our experiments demonstrate the effectiveness of the proposed method on speech, music, and environmental signals.
\end{abstract}


\section{Introduction}

Phase reconstruction of short-time Fourier transform (STFT) coefficients has been studied for decades~\cite{Griffin1984,Perraudin2013,Masuyama2019a,Peer2022,Prusa2017,Takamichi2018}.
When the STFT magnitude is synthesized or processed in the time--frequency domain, the corresponding phase is required to convert it to the time domain by using the inverse STFT.
While various phase reconstruction methods have been developed~\cite{Prusa2017,Takamichi2018}, the Griffin--Lim algorithm (GLA)~\cite{Griffin1984} has been widely used because it works without any assumptions on the target signal.
GLA leverages the redundancy of STFT and reconstructs the phase based on the consistency between complex STFT coefficients and time-domain signals.
Recently, variants of GLA have been developed from the point of view of optimization algorithms~\cite{Perraudin2013,Masuyama2019a,Peer2022} and through integration with deep neural networks~\cite{Masuyama2019b,Masuyama2021}.
These methods have shown promising performance when the given magnitude does not contain errors.

Recently, text-to-speech~\cite{Shen2018,Ping2018,Ren2021} and voice conversion~\cite{Kaneko2020,Hayashi2021} pipelines predict a low-dimensional spectral representation, such as a mel-spectrogram, and reconstruct a time-domain signal from the phaseless representation.
The auditory-motivated representation efficiently preserves the essential information and is easier to predict than the full-band magnitude.
The second stage of the pipelines requires mel-spectrogram inversion that reconstructs a time-domain signal from the given mel-spectrogram.
While neural vocoders directly reconstruct a time-domain signal~\cite{Prenger2019,Kumar2019,Kong2020}, recent studies reconstruct complex STFT coefficients and incorporate the inverse STFT~\cite{Kaneko2022,Webber2022} to reduce computational complexity.
These studies are relevant to phase reconstruction, and phase reconstruction is still used for synthesizing more general audio signals~\cite{Okamoto2022,Giorgi2022}.

\begin{figure}[t!]
\centering
\includegraphics[width=0.99\columnwidth]{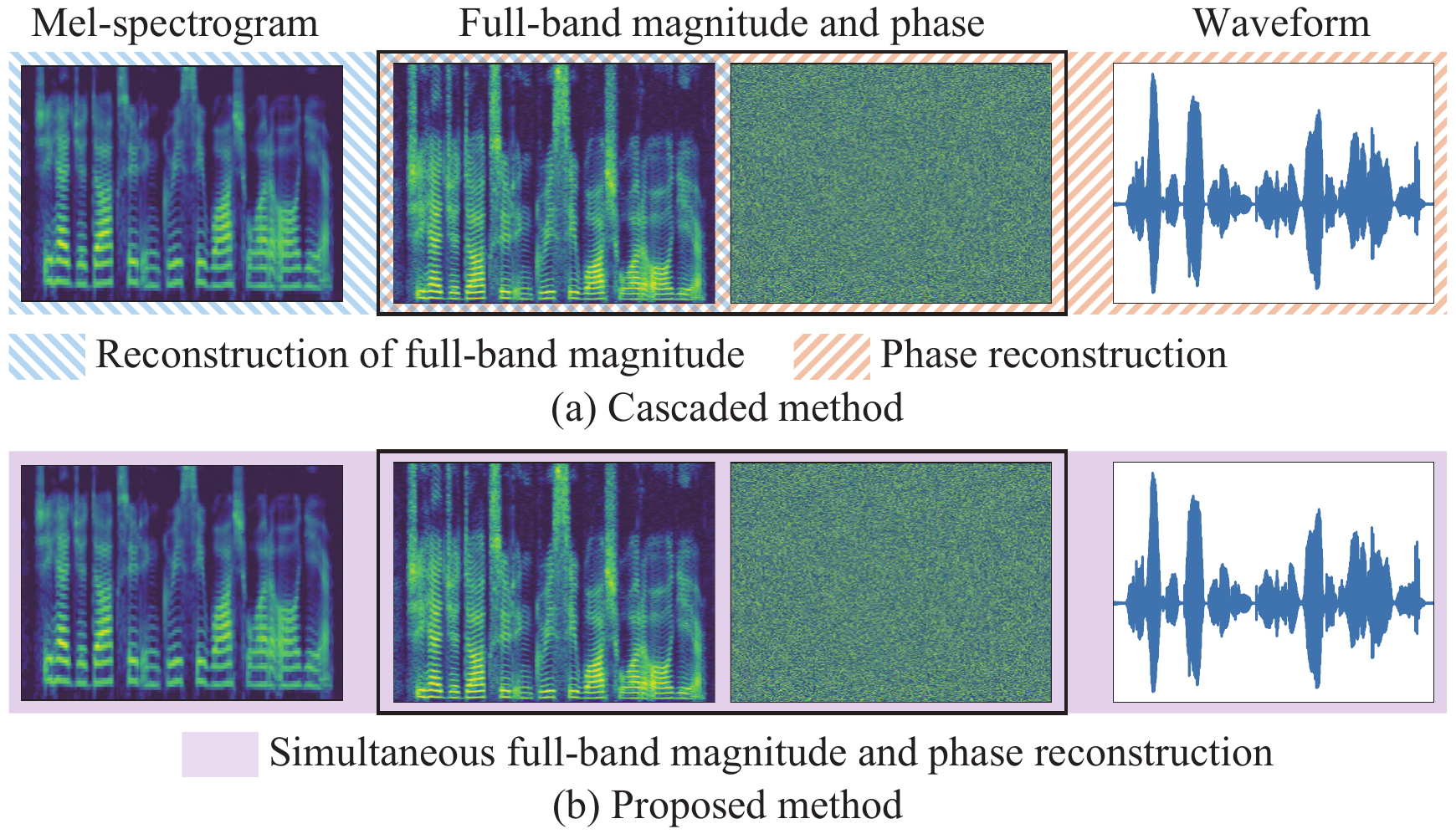}
\vspace{-4pt}
\caption{Illustrations of the (a) cascaded method and (b) simultaneous optimization method for reconstructing the audio signal from a given mel-spectrogram.
}
\label{fig:overview}
\vspace{-4pt}
\end{figure}

When we use phase reconstruction in mel-spectrogram inversion, we first reconstruct the full-band magnitude and then estimate the phase from the reconstructed full-band magnitude as illustrated in Fig.~\ref{fig:overview} (a).
We call such a two-stage method the cascaded method.
While the non-negative least squares method has been widely used to reconstruct the full-band magnitude, it reconstructs the magnitude frame-by-frame and cannot obtain the original full-band magnitude in common settings.
The reconstruction error deteriorates the performance of the cascaded method because the phase is estimated to be consistent with the reconstructed full-band magnitude.

To address this problem, we explore optimization-based methods that jointly reconstruct the full-band magnitude and phase as depicted in Fig.~\ref{fig:overview} (b).
We integrate the optimization problems for full-band magnitude reconstruction and phase reconstruction into a single problem.
We further modify its cost function, which makes the method less sensitive to a hyperparameter.
To solve the problems, we adopt the inertial proximal alternating linearized minimization (iPALM)~\cite{Pock2016} that can handle non-convex and non-smooth functions.
In our experiments with speech signals, the proposed methods outperformed the cascaded method in terms of PESQ~\cite{wpesq} and the extended STOI (ESTOI)~\cite{Jensen2016}.
We also demonstrated the advantage of the proposed method on music signals and foley sounds%
\footnote{
\href{https://yoshikimas.github.io/signal-reconstruction-from-mel-spectrogram}{yoshikimas.github.io/signal-reconstruction-from-mel-spectrogram}%
}.

\section{Preliminaries}

\subsection{Phase Reconstruction with Original-scale Magnitude}

Let the STFT coefficients of an audio signal $\bm{x}$ be $\bm{X} = \mathcal{G}(\bm{x}) \in \mathbb{C}^{F \times T}$, where $F$ and $T$ are the number of frequency bins and time frames, respectively.
As STFT is a redundant transform in common settings, the image of STFT $\mathcal{C}$ is a linear subspace of $\mathbb{C}^{F \times T}$.
If complex STFT coefficients $\bm{X}$ are not in $\mathcal{C}$, the magnitude of the reconstructed signal differs from the original one as follows:
\begin{equation}
|\mathcal{G} (\mathcal{G}^\dagger(\bm{X}))|
\neq |\bm{X}|,
\label{eq:inconsistent}
\end{equation}
where $(\cdot)^\dagger$ denotes the Moore-Penrose pseudo-inverse, and $|\cdot|$ denotes entry-wise absolute value.
Note that $\mathcal{G}^\dagger(\cdot)$ is the popular inverse STFT.
The discrepancy in \eqref{eq:inconsistent} is adverse in many audio applications since the magnitude of $\bm{X}$ is processed to the desired one.

To tackle this problem, phase reconstruction, or spectrogram inversion, is formulated as the following problem~\cite{Masuyama2019a}:
\begin{equation}
\text{Find} \hspace{5pt} \bm{X} \hspace{5pt} \text{s.t.} \hspace{5pt} {\bm{X} \in \mathcal{C} \cap \mathcal{A}},
\label{eq:union}
\end{equation}
where $\mathcal{A}$ is a set of STFT coefficients whose magnitude is equal to the given one $\bm{A} \in \mathbb{R}_+^{F \times T}$:
\begin{equation}
    \mathcal{A} = \{ \bm{X} \in \mathbb{C}^{F \times T}
    \mid |\bm{X}| = \bm{A} \} \hspace{3pt}.
\end{equation}
When $\bm{A}$ is artificially generated, $\mathcal{C} \cap \mathcal{A}$ can be empty.
In such cases, there is no solution for \eqref{eq:union}, and we can consider the following optimization problem instead:
\begin{equation}
    \min_{\bm{X} \in \mathcal{C}} \hspace{3pt} \mathcal{L}(\bm{X}, \bm{A}) = \lVert |\bm{X}| - \bm{A} \rVert^2,
    \label{eq:gla-opt}
\end{equation}
where $\lVert \cdot \rVert$ denotes the Frobenius norm.
The magnitude of the reconstructed signal $\mathcal{G}^\dagger(\bm{X})$ keeps the given one $\bm{A}$ as much as possible in terms of the squared Frobenius norm.

\subsection{Griffin--Lim Algorithm (GLA)}

GLA reconstructs the phase based on the redundancy of STFT and is implemented as alternating projections onto the sets in \eqref{eq:union}~\cite{Griffin1984}:
\begin{equation}
    \bm{X}^{[k+1]} = P_{\mathcal{C}} \left(P_{\mathcal{A}} \left(\bm{X}^{[k]}\right)\right),
    \label{eq:gla}
\end{equation}
where $k$ is the iteration index, $P_\mathcal{S}(\cdot)$ is the projection onto a set $\mathcal{S}$:
\begin{equation}
    P_{\mathcal{S}}(\bm{Y}) = \argmin_{\bm{X}} \iota_\mathcal{S}(\bm{X}) + \lVert \bm{X} - \bm{Y} \rVert^2,
    \label{eq:def-proj}
\end{equation}
and the indicator function $\iota_\mathcal{S}(\cdot)$ is defined as
\begin{equation}
    \iota_\mathcal{S}(\bm{X}) = 
    \begin{cases}
    0 & (\bm{X} \in \mathcal{S}) \\
    \infty & (\bm{X} \notin \mathcal{S})
    \end{cases}.
\end{equation}
The projection onto the set $\mathcal{C}$ is given by
\begin{equation}
    P_\mathcal{C}(\bm{X}) = \mathcal{G} (\mathcal{G}^\dagger(\bm{X})).
\end{equation}
Meanwhile, the projection onto the set $\mathcal{A}$ is defined as
\begin{equation}
P_{\mathcal{A}}(\bm{X}) = \bm{A} \odot \bm{X} \oslash |\bm{X}|,
\label{eq:proj-A}
\end{equation}
where $\odot$ and $\oslash$ are entry-wise multiplication and division, respectively, and we set $0/|0| = 0$ as in \cite{Masuyama2019a,Peer2022}.

We can derive the iterative procedure of GLA in \eqref{eq:gla} by applying the projected gradient method to \eqref{eq:gla-opt}~\cite{Zhang2017,Chen2018}%
\footnote{
The cost function in \eqref{eq:gla-opt} has non-smooth points, where we set the gradient to zero and obtain $\nabla_{\bm{X}} \mathcal{L}(\bm{X}, \bm{A})$ in \eqref{eq:gradient}.
With abuse of terminology, we still refer to it as the gradient for simplicity as in \cite{Zhang2017}.
}%
:
\begin{equation}
    \bm{X}^{[k+1]} = P_{\mathcal{C}}(\bm{X}^{[k]} - \mu \nabla_{\bm{X}} \mathcal{L}(\bm{X}^{[k]}, \bm{A})),
    \label{eq:gla-pg}
\end{equation}
where $\mu \in \mathbb{R}_+$ is the step size, and the gradient for $\mathcal{L}(\bm{X}, \bm{A})$ is given by
\begin{equation}
    \nabla_{\bm{X}} \mathcal{L}(\bm{X}, \bm{A}) = \bm{X} - P_{\mathcal{A}}(\bm{X}).
    \label{eq:gradient}
\end{equation}
By substituting \eqref{eq:gradient} to \eqref{eq:gla-pg} and setting $\mu$ to $1$, the projected gradient method coincides with the alternating projections in \eqref{eq:gla}.
This interpretation of GLA is more relevant to the proposed method.

\subsection{Mel-Spectrogram Inversion}
\label{sec:cascade}

In recent text-to-speech pipelines~\cite{Shen2018,Ping2018,Ren2021}, a mel-spectrogram has been used as an intermediate representation.
Mel-spectrogram inversion aims to reconstruct a time-domain signal from a mel-spectrogram $\bm{M} \in \mathbb{R}_+^{B \times T}$ given by $\bm{E} \bm{A}$, where $B$ is the number of mel bins, and $\bm{E} \in \mathbb{R}^{B \times F}$ converts the frequency scale.

Since phase reconstruction has been successfully addressed, a simple method for mel-spectrogram inversion is to reconstruct the full-band magnitude in advance and then apply a phase reconstruction method.
The reconstruction of the full-band magnitude can be formulated as follows:
\begin{subequations}%
\begin{align}
    \min_{\bm{Y}} &\hspace{5pt} \frac{1}{2} \lVert \bm{E} \bm{Y} - \bm{M} \rVert^2, \\
    \text{s.t.} \, &\hspace{5pt} \bm{Y} \in \mathcal{N},
\end{align}%
\label{eq:mel-to-full}%
\end{subequations}%
where $\mathcal{N} = \mathbb{R}_+^{F \times T}$.
The cascaded method for mel-spectrogram inversion with \eqref{eq:mel-to-full} has been used in popular audio analysis packages including \texttt{Librosa}~\cite{McFee2015}.
The optimization problem in \eqref{eq:mel-to-full} did not consider the relationship between adjacent time frames in STFT, and its solution is not unique in common settings.
As a result, the reconstructed full-band magnitude $\bm{Y}$ cannot be consistent with a time-domain signal, which deteriorates the performance of the subsequent phase reconstruction.

\section{Proposed Mel-Spectrogram Inversion}

In this section, we present two mel-spectrogram inversion methods that estimate full-band magnitude and phase jointly.

\subsection{Joint Full-band Magnitude and Phase Reconstruction}
\label{sec:naive}

As discussed in Section~\ref{sec:cascade}, the cascaded mel-spectrogram inversion method reconstructs the full-band magnitude frame-by-frame and results in sub-optimal performance due to the inconsistency in the reconstructed full-band magnitude.
Meanwhile, we can reduce the inconsistency by leveraging the redundancy of STFT such as the signal overlaps in adjacent time frames.
We thus simultaneously consider the bi-level relationships:
the relationship between a mel-spectrogram and a full-band
magnitude; and the relationship between full-band STFT coefficients and a time-domain signal as shown in Fig.~\ref{fig:overview} (b).
The former and latter relationships are induced by $\bm{M} = \bm{E}\bm{A}$ and STFT, respectively.

In detail, to estimate the full-band magnitude and phase jointly, we integrate the optimization problems in \eqref{eq:gla-opt} and \eqref{eq:mel-to-full} as follows:
\begin{subequations}%
\begin{align}
\min_{\bm{X}, \bm{Y}} &\hspace{3pt} \frac{1}{2} \mathcal{L}(\bm{X}, \bm{Y}) + \frac{\lambda}{2} \lVert \bm{E} \bm{Y} - \bm{M} \rVert^2, \label{eq:prop-combination-cost} \\
\text{s.t.} \hspace{3pt} &\hspace{3pt} \bm{X} \in \mathcal{C},
\hspace{3pt} \bm{Y} \in \mathcal{N},
\end{align}
\label{eq:prop-combination}%
\end{subequations}%
where $\lambda \in \mathbb{R}_+$ is a hyperparamter.
The reconstruction of the full-band magnitude considers its bi-level relationships with the mel-spectrogram and with the time-domain signal.

As another formulation, we can replace the first term in \eqref{eq:prop-combination-cost} to a constraint: $\bm{Y} = |\bm{X}|$.
The obtained optimization problem can be reformulated as the following unconstrained optimization problem:
\begin{equation}
    \min_{\bm{x}} \hspace{3pt} \frac{1}{2} \lVert \bm{E} |\mathcal{G}(\bm{x})| - \bm{M} \rVert^2.
    \label{eq:prop-waveform}
\end{equation}
While we can apply the gradient descent method to \eqref{eq:prop-waveform} as in GLA, we will demonstrate that applying the iPALM algorithm for \eqref{eq:prop-combination} improves the quality of the reconstructed signal faster in Section~\ref{sec:speech-ex}.

\subsection{Squared Distance to Set as Mel-spectrogram Fidelity}
\label{sec:prop}

In \eqref{eq:prop-combination-cost}, the two cost functions are defined on different domains: the full-band magnitude and the mel-spectrogram, which makes it difficult to tune $\lambda$ intuitively. 
We thus replace the second term in \eqref{eq:prop-combination-cost} by the squared distance to the following set $\mathcal{M}$:
\begin{align}
    \frac{1}{2} d_{\mathcal{M}}^2(\bm{Y}) &= \inf_{\bm{Z} \in \mathcal{M}} \frac{1}{2} \lVert \bm{Y} -\bm {Z} \rVert^2,
    \label{eq:dist-to-setM} \\
    \mathcal{M} &= \{ \bm{Z} \in \mathbb{R}^{F \times T} \mid \bm{E} \bm{Z} = \bm{M} \},
\end{align}
This cost function is computed on the full-band magnitude as the first term of \eqref{eq:prop-combination-cost}.
The gradient for \eqref{eq:dist-to-setM} is given by~\cite{Parikh2014}
\begin{align}
    \nabla \frac{1}{2} d_{\mathcal{M}}^2(\bm{Y}) &= \bm{Y} - P_{\mathcal{M}}(\bm{Y}), \\
    P_{\mathcal{M}}(\bm{Y}) &= \bm{Y} - \bm{E}^\dagger(\bm{E}\bm{Y} -\bm{M}).
\end{align}
By using the cost function in \eqref{eq:dist-to-setM}, we formulate another optimization problem for mel-specrogram inversion:
\begin{subequations}%
\begin{align}
\min_{\bm{X}, \bm{Y}} &\hspace{3pt} \frac{1}{2} \mathcal{L}(\bm{X}, \bm{Y}) + \frac{\lambda}{2} d_{\mathcal{M}}^2(\bm{Y}),  \\
\text{s.t.} \hspace{3pt} &\hspace{3pt} \bm{X} \in \mathcal{C},
\hspace{3pt} \bm{Y} \in \mathcal{N}.
\end{align}
\label{eq:prop-problem}%
\end{subequations}%
We will demonstrate that \eqref{eq:prop-problem} is relatively insensitive to the value of $\lambda$ than \eqref{eq:prop-combination} in Section~\ref{sec:speech-ex}.

\subsection{iPALM Algorithm for \texorpdfstring{\eqref{eq:prop-problem}}{(19)}}
\label{sec:ipalm}

To solve the optimization problem in \eqref{eq:prop-problem}, in this paper, we adopt the iPALM algorithm~\cite{Pock2016}.
Let us consider an optimization problem with three cost functions:
\begin{equation}
    \min_{\bm{X}, \bm{Y}} \mathcal{F}_1(\bm{X}) + \mathcal{F}_2(\bm{Y}) + \mathcal{H}(\bm{X}, \bm{Y}),
    \label{eq:general-problem}
\end{equation}
where the cost functions can be non-convex.
We assume the following proximity operator for $\mathcal{F}_1(\cdot)$ and $\mathcal{F}_2(\cdot)$ is calculated efficiently:
\begin{equation}
    \mathrm{prox}_{\mathcal{F}}(\bm{X}) = \argmin_{\bm{Z}} \mathcal{F}(\bm{Z}) + \frac{1}{2} \lVert  \bm{Z} - \bm{X} \rVert^2.
\end{equation}
The iPALM algorithm solves the optimization problem in \eqref{eq:general-problem} by iterating the following procedure:
\begin{subequations}%
\begin{align}
\!\!\!\! \underline{\bm{X}}^{[k]} &= \bm{X}^{[k]} + \alpha (\bm{X}^{[k]} - \bm{X}^{[k-1]}), \\
\!\!\!\! \bm{X}^{[k+1]} &=  \mathrm{prox}_{\mathcal{F}_1/\tau_1^{[k]}} \!\! \left( \!
\underline{\bm{X}}^{[k]} - \frac{1}{\tau_1^{[k]}} \nabla_{\bm{X}} \mathcal{H}(\underline{\bm{X}}^{[k]}, \bm{Y}^{[k]}) \!
\right) \!\! , \!\! \label{eq:xupdate} \\
\!\!\!\! \underline{\bm{Y}}^{[k]} &= \bm{Y}^{[k]} + \beta (\bm{Y}^{[k]} - \bm{Y}^{[k-1]}), \\
\!\!\!\! \bm{Y}^{[k+1]} &= \mathrm{prox}_{\mathcal{F}_2/\tau_2^{[k]}} \!\! \left( \!
\underline{\bm{Y}}^{[k]} - \frac{1}{\tau_2^{[k]}} \nabla_{\bm{Y}} \mathcal{H}(\bm{X}^{[k+1]}, \underline{\bm{Y}}^{[k]}) \!
\right) \!\! , \!\!\!\! \label{eq:yupdate}
\end{align}
\label{eq:ipalm}%
\end{subequations}%
where $\alpha \in \mathbb{R}_+$ and $\beta \in \mathbb{R}_+$ are the inertial parameters.
When both inertial parameters are zero, the iterative procedure coincides with the original PALM algorithm~\cite{Bolte2013}.

To use the iPALM algorithm for \eqref{eq:prop-problem}, we replace its constraints with indicator functions and reformulate it into the form of \eqref{eq:general-problem}:
\begin{equation}
\min_{\bm{X}, \bm{Y}} \hspace{3pt} \iota_{\mathcal{C}}(\bm{X}) + \iota_{\mathcal{N}}(\bm{Y}) + \mathcal{I}(\bm{X}, \bm{Y}),
\label{eq:reformulated}
\end{equation}
where $\mathcal{I}(\bm{X}, \bm{Y})$ is given by
\begin{equation}
\mathcal{I}(\bm{X}, \bm{Y}) = \frac{1}{2} \mathcal{L}(\bm{X}, \bm{Y}) + \frac{\lambda}{2} d_{\mathcal{M}}^2(\bm{Y}).
\end{equation}
By substituting each term of \eqref{eq:reformulated} into \eqref{eq:ipalm}, we yield Algorithm~\ref{alg:prop}, where we set $\beta$ to $0$ and $\tau_1^{[k]}$ to $1/2$ for simplicity%
\footnote{
To guarantee the convergence to a critical point, $\tau_1^{[k]}$ should be tuned in each iteration~\cite{Pock2016}.
However, we experimentally confirmed that Algorithm~\ref{alg:prop} stably works even with $\tau_1^{[k]} = 1/2$ regardless of the iteration index $k$.
}%
.
Meanwhile, we fix $\tau_2^{[k]}$ to $1 + \lambda$ based on the partial Lipschitz constant of the gradient of $\mathcal{I}(\cdot, \cdot)$.
The proximity operator of each indicator function is the projection by definition, and the projection onto the set $\mathcal{N}$ is computed by taking the maximum with zero entry-wise.
According to \eqref{eq:xupdate}, the update of $\bm{X}^{[k]}$ corresponds to the projected gradient method for minimizing $\mathcal{I}(\bm{X}, \bm{Y}^{[k]})$ on $\mathcal{C}$.
By extending the interpretation of GLA as the projected gradient method in \eqref{eq:gla-pg}, we obtain the update as $P_{\mathcal{C}}
(P_{\mathcal{Y}^{[k]}} (\underline{\bm{X}}^{[k]}))$, where $\mathcal{Y}^{[k]}$ is the set of the STFT coefficients whose magnitude is equal to $\bm{Y}^{[k]}$ at all time--frequency bins.
More precisely, the update of $\bm{X}$ with the inertial acceleration corresponds to the fast GLA (FGLA)~\cite{Perraudin2013}.

The iPALM algorithm for \eqref{eq:prop-combination} requires changing the gradient in \eqref{eq:yupdate} from that for \eqref{eq:prop-problem}.
As a result, the update of $\bm{Z}$ in Algorithm~\ref{alg:prop} is replaced as follows:
\begin{equation}
    \bm{Z}^{[k]} = \bm{Y}^{[k]} - \frac{1}{1+\lambda} \left( \bm{Y}^{[k]} - |\bm{X}^{[k+1]}| + \lambda \bm{G} \bm{Y}^{[k]} - \lambda \bm{v} \right), \!\!
\end{equation}
where $\bm{G} = \bm{E}^\mathsf{T} \bm{E}$, $\bm{v} = \bm{E}^\mathsf{T} \bm{M}$, and $(\cdot)^\mathsf{T}$ denotes the transpose.

\begin{algorithm}[t!]
\caption{iPALM for mel-pectrogram inversion}
\algsetup{indent=2mm}
\begin{algorithmic}
\renewcommand{\algorithmicrequire}{\textbf{Input:}}
\renewcommand{\algorithmicensure}{\textbf{Output:}}
\REQUIRE $\bm{X}^{[-1]}$, $\bm{X}^{[0]}$, $\bm{Y}^{[0]}$, $\lambda$, $\alpha$
\ENSURE $\bm{X}^{[K]}$
\FOR {$k=0, \ldots, K-1$}
\STATE $\underline{\bm{X}}^{[k]} = \bm{X}^{[k]} + \alpha (\bm{X}^{[k]} - \bm{X}^{[k-1]})$
\STATE ${\bm{X}^{[k+1]}} = {P_{\mathcal{C}}}
(P_{\mathcal{Y}^{[k]}}(\underline{\bm{X}}^{[k]}))$
\STATE $\bm{Z}^{[k]} = 1 / (1+\lambda) |\bm{X}^{[k+1]}| + \lambda / (1+\lambda) P_{\mathcal{\mathcal{M}}}(\bm{Y}^{[k]})$
\STATE $\bm{Y}^{[k+1]} = P_{\mathcal{N}}(\bm{Z}^{[k]})$
\ENDFOR
\end{algorithmic}
\label{alg:prop}
\end{algorithm}

\section{Experiments}

\begin{figure}[t!]
\centering
\includegraphics[width=0.99\columnwidth]{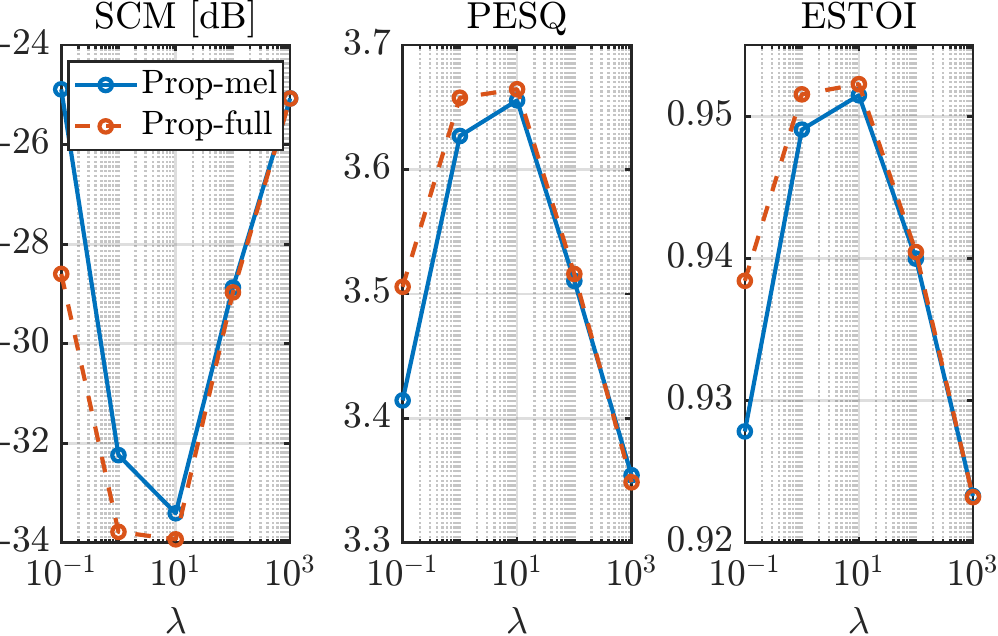}
\vspace{-6pt}
\caption{Average SCM/PESQ/ESTOI of the signals reconstructed by the proposed methods with different $\lambda$.
}
\label{fig:ex1_prop1_vs_prop2}
\end{figure}

\begin{figure}[t]
\centering
\vspace{-4pt}
\includegraphics[width=0.99\columnwidth]{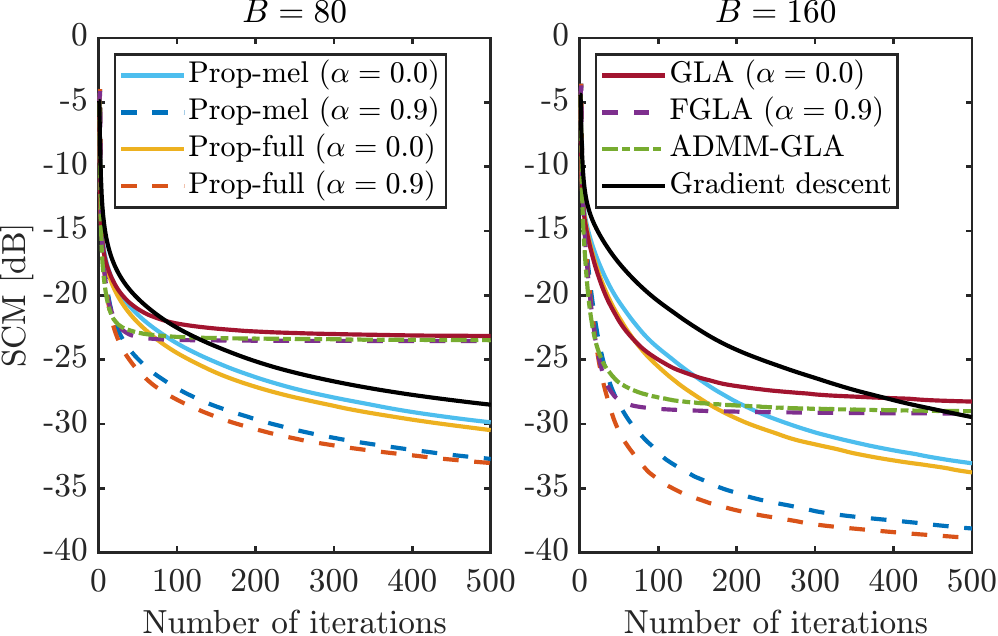}
\vspace{-4pt}
\caption{
SCM with respect to the number of iteration.
}
\label{fig:scm_per_iter}
\vspace{-6pt}
\end{figure}

\subsection{Evaluation on Speech Signals}
\label{sec:speech-ex}

In this section, we demonstrate the effectiveness of the proposed methods on speech signals.
We used $200$ utterances of the TIMIT dataset provided in \cite{Mowlaee2016a}, where the half was for tuning $\lambda$ and the rest was for evaluation.
The sampling rate was 16 kHz.
STFT was performed using the Hann window of $1024$ with a $256$-sample shift, and the number of mel bins was $80$, which is popular in recent text-to-speech pipelines~\cite{Shen2018,Ping2018,Ren2021}.
We first investigated the effect of $\lambda$ on the iPALM algorithms for \eqref{eq:prop-combination} and \eqref{eq:prop-problem}, where the algorithms are abbreviated as Prop-mel and Prop-full, respectively.
The number of iterations was $500$, and $\alpha=0.9$.
The reconstructed signals were evaluated by PESQ~\cite{wpesq}, ESTOI~\cite{Jensen2016}, and the spectral convergence on mel-spectrogram:
\begin{equation}
    \text{SCM} = 20 \log_{10} \left( \frac{ \lVert \bm{E} |\mathcal{G}(\widehat{\bm{x}})| - \bm{M} \rVert}{\lVert \bm{M} \rVert} \right),
\end{equation}
where $\widehat{\bm{x}}$ is the reconstructed signal.
The objective measures with different $\lambda$ are shown in Fig.~\ref{fig:ex1_prop1_vs_prop2}.
Both methods performed best with $\lambda=10$, and thus we used this value in the following experiments.
Prop-full is relatively insensitive to the value of $\lambda$, which will make the hyperparameter tuning easier in other conditions.

Next, we compared the proposed methods with the cascaded method, where \texttt{mel\_to\_stft} in \texttt{Librosa} solved the optimization problem in \eqref{eq:mel-to-full} by the L-BFGS-B algorithm~\cite{Byrd1995}.
Phase reconstruction was performed by GLA~\cite{Griffin1984}, FGLA~\cite{Perraudin2013}, and ADMM-GLA~\cite{Masuyama2019a}.
We also evaluated the gradient descent method for \eqref{eq:prop-waveform}.
Fig.~\ref{fig:scm_per_iter} shows SCM for various methods per iteration.
Regardless of the phase reconstruction methods, the performance of the cascaded method was limited due to the error in the intermediate reconstruction of the full-band magnitude, especially when $B = 80$.
The performance improvement of the gradient descent method was slower than the iPALM algorithms with $\alpha=0.0$, while the computational complexity is the same.
The inertial acceleration further improved the performance.
Table~\ref{tab:pesq_estoi} shows the PESQ and ESTOI after $500$ iterations, and Prop-full with $\alpha=0.9$ performed best.

\begin{table}[t!]
    \centering
    \caption{PESQ and ESTOI with different number of mel bins.}
    \scalebox{1.}[1.]{
    \begin{tabular}{l|cc|cc}
        \toprule
         & \multicolumn{2}{c|}{$B = 80$} & \multicolumn{2}{c}{$B=160$} \\
        \cmidrule{2-5}
         & PESQ & ESTOI & PESQ & ESTOI \\
        \midrule
        GLA ($\alpha = 0.0$)  & 3.31 & 0.91 & 3.82 & 0.96 \\
        FGLA ($\alpha = 0.9$) & 3.30 & 0.92 & 3.89 & 0.97 \\
        \midrule
        Gradient descent & 3.57 & 0.94 & 3.86 & 0.97 \\
        Prop-mel ($\alpha = 0.0$) & 3.60 & 0.94 & 3.92 & 0.97 \\
        Prop-mel ($\alpha = 0.9$) & 3.65 & \bf{0.95} & 4.05 & \bf{0.98} \\
        Prop-full ($\alpha = 0.0$) & 3.61 & \bf{0.95} & 3.94 & 0.97 \\
        Prop-full ($\alpha = 0.9$) & \bf{3.68} & \bf{0.95} & \bf{4.06} & \bf{0.98} \\
        \bottomrule
    \end{tabular}
    }
    \label{tab:pesq_estoi}
\end{table}

\vspace{-6pt}
\subsection{Evaluation on Music and Environmental Signals}
\label{sec:other-ex}

To demonstrate the effectiveness of the joint optimization, we compared Prop-full with the cascaded method using FGLA.
As music signals, 12 song snippets from the MASS dataset%
\footnote{\href{https://www.upf.edu/web/mtg/mass}{www.upf.edu/web/mtg/mass}}
were used.
The sampling rate was $44100$ Hz, and mel-spectrograms with $96$ bins were computed with the Hann window of $2048$ samples and the $256$-sample shift \cite{Giorgi2022}.
We also investigated the performance on foley sounds from the development sets of DCASE2023 Task 7%
\footnote{\href{https://dcase.community/challenge2023/task-foley-sound-synthesis}{dcase.community/challenge2023/task-foley-sound-synthesis}}%
.
We used 50 signals for each of the classes of foley sounds.
While the sampling rate was $22050$ Hz, the STFT conditions were the same as in Section~\ref{sec:speech-ex}, which coincides with the challenge baseline.

SCMs for Prop-full and the cascaded method using FGLA are shown in Fig.~\ref{fig:scatter}.
While the performance of both methods was diverse, Prop-full achieved better performance regardless of the sound class.
This result confirms the effectiveness of the proposed method not only for speech but also for general audio signals.

\begin{figure}[t!]
\centering
\vspace{-6pt}
\includegraphics[width=0.85\columnwidth]{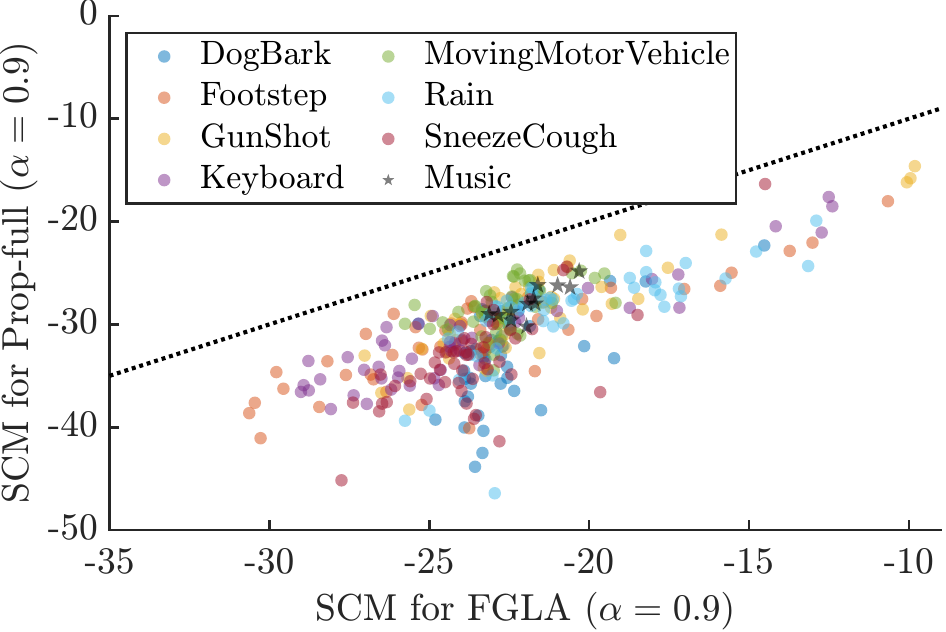}
\vspace{-4pt}
\caption{
SCM for Prop-full versus SCM for the cascaded method using FGLA.
The dotted line indicates equal performance.
}
\label{fig:scatter}
\end{figure}

\section{Conclusion}
\vspace{-4pt}

In this paper, we propose to jointly estimate the full-band magnitude and phase based on the bi-level consistency.
We explore two optimization problems and derive the iPALM algorithms for them.
Our experimental results on speech, music, and environmental signals show the advantage of the proposed method for general audio signals.
We will improve the optimization algorithms to reduce the number of iterations for obtaining a signal with sufficient quality.

\vspace{-4pt}
\section{Acknowledgment}
\vspace{-4pt}
This work was supported by JSPS KAKENHI Grant Numbers JP20H00613 and JP21J21371, and JST CREST Grant Number JPMJCR19A3, Japan.

\bibliographystyle{IEEEtran}
\setlength{\itemsep}{-5pt}
\bibliography{references}

\end{sloppy}
\end{document}